# LARGE THERMOPOWER IN METALLIC OXIDES: MISFIT COBALTITES AND MANGANO-RUTHENATES

S. Hébert, C. Martin, A. Maignan, R. Frésard, J. Hejtmanek* and B. Raveau

Laboratoire CRISMAT, UMR 6508 associée au CNRS et ISMRA,
6 Bd du Maréchal Juin, 14050 Caen Cedex, France
(sylvie.hebert@ismra.fr)
*Institute of Physics, Academy of Sciences of the Czech Republic, Cukrovarnicka 10, 16200 Praha 6, Czech Republic

**ABSTRACT**

Two different kinds of metal transition oxides have been studied for their large thermopower values. The first one corresponds to the Tl-based misfit cobaltite which is a hole-doped metal. We demonstrate that the partial Bi-substitution for Tl in this phase induces an increase of the room temperature (RT) thermopower (TEP) value. Same result is obtained with the new $Pb_{1/3}SrCoO_{3+\delta}$ misfit corresponding to the Tl complete replacement by lead. Simultaneously, the T dependence of their resistivity exhibits a re-entrance below 70-90K where a large negative magnetoresistance is observed. Magnetic measurements reveal a strong interplay between spins and charges for this class of materials.

Electron-doped (n-type) perovskite manganites are a second class of potential candidates for applications. In particular, the $Ru^{4+/5+}$ substitution for Mn in the $CaMnO_3$ semi-conductor induces a drastic drop of the resistivity values. Metals with large RT TEP values and not too large thermal conductivities are generated. A comparison with best known materials, $Bi_2Te_3$ and $NaCo_2O_4$ is made.

## 1. INTRODUCTION

Metal transition oxides exhibit various kinds of spectacular physical properties such as high $T_C$ superconductivity (HTCS) in cuprates and colossal magneto-resistance in manganites. More recently, Terasaki et al [1] have shown that the $NaCo_2O_4$ cobaltite exhibits a high figure of merit Z and since then the research for new metallic oxides with large thermopower (TEP) values to be used for applications to convert heat in energy has become an important topic. For instance good characteristics (large TEP and low resistivity) have been obtained for $Sr_{1-x}La_xTiO_3$ titanates [2] and the fabrication of an all-oxide thermoelectric power generator based on a 'misfit cobaltite' and a perovskite manganite has been very recently reported [3]. In the latter device, the p-type part is made of Gd-doped $Ca_3Co_4O_9$ misfit phase. The first misfit cobalt oxide, $Tl_{0.4}SrCoO_{3+\delta}$, was discovered in our laboratory [4] and the refinement of the misfit structure has been performed for the bismuth based oxide $[Bi_{0.87}SrO_2]_2[CoO_2]_{1.82}$ [5]. On the other hand, the n-type leg of the device described in ref. [3] consists in a $La_{1-x}Ca_xMnO_3$ perovskite with x ~ 0.82. The latter is an 'electron-doped' manganite since $Mn^{3+}$ holes in the $Mn^{4+}$ matrix are created by the lanthanum for calcium substitution [6]. Interestingly for TEP purposes, electron-doped manganites are better metals than the hole-doped ones due to their lower $Mn^{3+}$ concentration. This Jahn-Teller cation is at the origin of the polaronic character of the transport properties.

We have thus started recently the research of new potential candidates for TEP applications in two classes of oxides, misfit cobaltites and electron-doped perovskite manganites. In the following, our results concerning the magnetic and transport properties of the former and the latter are described in parts 2 and 3, respectively.

## 2. MISFIT COBALTITES

The structures of the $NaCo_2O_4$ [7] and the misfit cobaltites contain the same type of $CoO_2$ layers. The edge-shared $CoO_6$ octahedra form a layer of triangular cobalt in which the trivalent and tetravalent cobalt cations remain in the $t_{2g}^6$ and $t_{2g}^5$ low-spin state, respectively. The rhombohedral crystal field splits further the $t_{2g}$ orbitals in one $a_{1g}$ and two $e'_g$ orbitals with light and heavy holes, respectively [8]. These $CoO_2$ layers are metallic and they are thought to be responsible for the large thermopower TEP values. In contrast, the array of corner-shared $CoO_6$ octahedra in the $La_{1-x}Sr_xCoO_3$ perovskite exhibits only small TEP values when the $Co^{4+}$ concentration becomes sufficient to reach metallicity [9].

The structural difference between $NaCo_2O_4$ and misfit cobaltites lies thus in the separating layer, randomly half-filled Na layers and rock-salt type layers for the former and the latter, respectively. The misfit character comes from the structural discordance between their two kinds of monoclinic sublattices: the $CdI_2$-type $CoO_2$ and rock-salt type layers have similar a, c and β parameters but different b cell parameters [4-5]. Consequently, the transport properties are strongly anisotropic, metallic along the $CoO_2$ planes but quasi-insulating through the rock-salt type layers [11].

Among these misfit cobaltites, there exist two types of phases which differ by their c cell parameters (the c direction corresponds to the layer stacking). This reflects their different number of separating layers: c ~ 11Å and c ~ 15Å for 3 and 4 layers, respectively. The latter corresponds to the $[Bi_{0.82}SrO_2]_2[CoO_2]_{1.82}$ [5] misfit oxide and derived phases in which Pb replaces partially Bi. Their TEP properties have already been extensively reported see for instance ref. [10]. The other class of misfits corresponds to the Tl-based, $Tl_\alpha[SrO]_{1-x}[CoO_2]$ [4, 12], and CaCo-based $[Co_{0.5}CaO_{1.5-x}]_2[CoO_2]_{1.62}$ [11] phases, also written for sake of clarity $Tl_\alpha SrCoO_{3+\delta}$ and $Ca_3Co_4O_9$.

In the following we will focus on the Tl-based misfit cobaltite.

### 2.1 BISMUTH SUBSTITUTIONS FOR THALLIUM

Structural similarities between the RS type layers of misfit cobaltites and of HTCS have opened the route to the discovery of new Tl-based misfit cobaltites. Because bismuth and lead were substituted for Tl in the case of Tl-based cuprates such as the $TlSr_2CaCu_2O_7$ "1212" superconductor leading to the following formulae, $Tl_{2/3}Bi_{1/3}Sr_2CaCu_2O_7$ and $Tl_{1/2}Pb_{1/2}Sr_2CaCu_2O_7$, several attempts of Bi or Pb substitutions for Tl have been made in the Tl-based misfit prepared in silica tubes with a one-step method [4].

In the case of the bismuth substitution, it is found that a single misfit phase of analyzed cationic composition "$Tl_{0.28}Bi_{0.14}SrCo$" can be stabilized by starting from the nominal composition $(Tl_{2/3}Bi_{1/3})_{0.6}SrCoO_{4.28}$. It should be pointed out that by starting from the Tl/Bi = 1 ratio, a Tl/Bi = 2 final ratio is always measured by EDS, indicating the great stability of the phase corresponding to the ratio Tl/Bi = 2 as it is the case for the $Tl_{2/3}Bi_{1/3}$-1212 superconducting cuprate.

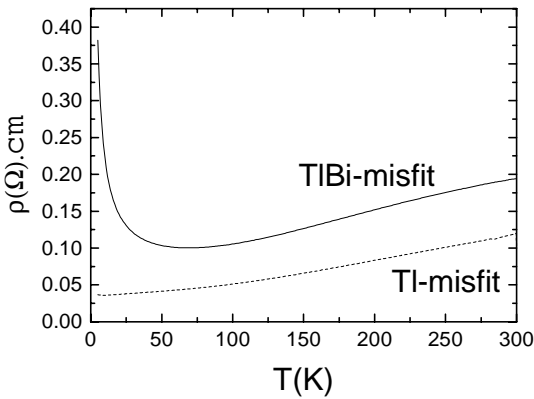

Fig. 1 : $\rho(T)$ curves of $Tl_{0.28}Bi_{0.14}SrCoO_{3+\delta}$ ('TlBi') and $Tl_{2/3}SrCoO_{3+\delta}$ ('Tl'). Tl and Bi contents are the results of EDS analyses.

Although the Bi for Tl partial substitution does not modify significantly the 300 K resistivity value, $\rho_{Tl/Bi}$ = 20 mΩ.cm against $\rho_{Tl}$ = 12 mΩ.cm for the more conductive sample (see the corresponding $\rho(T)$ curves in Fig. 1), the low-T part of the curve is dramatically changed by the Bi substitution which induces a semi-conducting like re-entrant behavior below 60 K. Accordingly, ρ reaches a minimum value of 10 mΩ.cm at 60 K and then increases as T further decreases to reach ρ = 38 mΩ.cm at 5 K. A similar ρ (T) shape was also found in the case of the $Ca_3Co_4O_9$ misfit [11] in the form of polycrystals or single-crystals (along the ab-plane). This shape modification of the ρ (T) curve compared with that of the Tl-Sr-Co-O misfits can be ascribed to the different magnetic behavior induced by Bi. As shown in the T dependent inverse susceptibility $\chi^{-1}(T)$ curves of Fig. 2, the TlBi misfit undergoes a magnetic transition below 10 K and $\chi^{-1}$ starts to deviate from the linearity below ~ 50 K, whereas the Tl-Sr-Co-O misfit remains paramagnetic down to 2K.

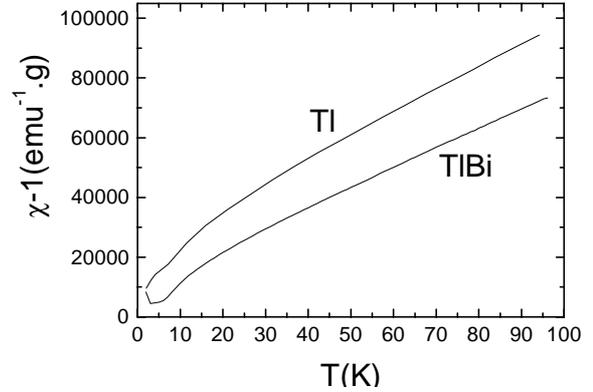

Fig. 2 : T dependent inverse magnetic susceptibility ($\chi^{-1}$) of "Tl" and "TlBi" misfits. $\chi^{-1}$ values are calculated from magnetization data measured with a SQUID magnetometer ($\mu_0 H$ = 0.3T).

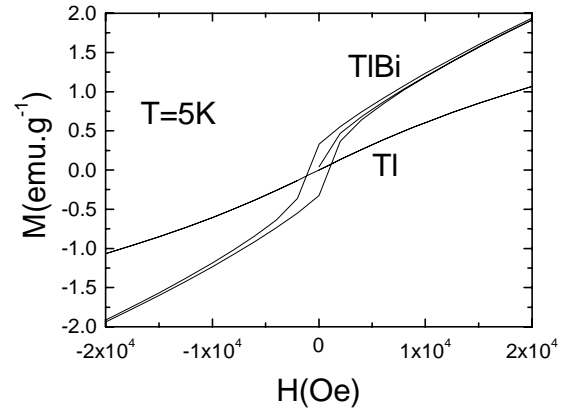

Fig. 3 : Magnetic field dependence of the magnetization at 5K for both "Tl" ($Tl_{0.4}SrCoO_{3+\delta}$) and "TlBi" ($Tl_{0.28}Bi_{0.14}SrCoO_{3+\delta}$) misfit cobaltites.

This is also shown in Fig. 3 where a clear hysteresis is found in the M(H) loop of the Tl/Bi misfit. Accordingly, in the absence of applied magnetic field, disorder in the spins alignment make that the resistivity increases as the sample becomes magnetic and under field application, the spin scattering reduction leads to a negative magnetoresistance (MR). If one defines MR as MR = -100 x [$\rho_H/\rho_{H=0}$-1], MR reaches 25 % in 7T at 2K for TlBi, whereas, in the absence of magnetic ordering, the magnetoresistance is positive in the Tl misfit reaching only 2% in 7T which is characteristic of a metal.

Interestingly, the thermopower values (Seebeck, S) of the TlBi-misfit are larger than those of the Tl-misfit (see Fig. 4), +120μV.K$^{-1}$ and +90μV.K$^{-1}$ at 300K for TlBi and Tl based misfit, respectively. In fact, the TlBi $S_{300 K}$ value is similar to that of $Ca_3Co_4O_9$ as also shown on this figure.

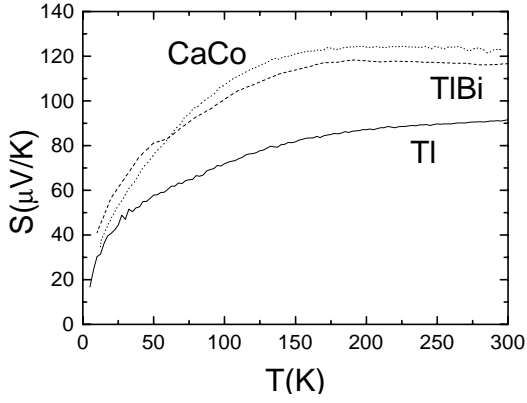

Fig. 4 : S(T) curves of "Tl"- and "TlBi"- and $Ca_3Co_4O_9$ "CaCo" misfit cobaltites.

## 2.2. LEAD SUBSTITUTION: $Pb_{1/3}SrCoO_{3+\delta}$ A NEW MISFIT COBALTITE

Similarly to the Bi substitution, we have tried to replace partially Tl by Pb. First of all, by varying the nominal Pb content from x = 0.1 to x = 0.3 in $Tl_{0.6-x}Pb_xSrCoO_{4.28}$, the EDS analyses of the reacted materials always reveal the existence of a phase mixture made of a Pb rich and a Pb poor misfit cobaltites, contrasting with the aforementioned Bi substitution. This result could be possibly due to the higher thermodynamical stability of the end members –pure Tl and pure Pb misfit- compared to the mixed Tl/Pb-misfit.

This has motivated us to try the synthesis of a pure Pb misfit cobaltite by using the following experimental: stoichiometric amounts of oxides and peroxides, $PbO_2$, $SrO_2$, $Co_3O_4$ are mixed and pressed in the form of bars. The latter are sealed under vacuum in silica tubes with approximately 0.8g of bars for a volume of $3cm^3$ in the tube. Then, the tubes are heated in 6h at 900°C for 12h and finally cooled down to RT in 6h. After the reaction completion, black ceramic bars are obtained.

Best samples quality is achieved by starting from $Pb_{0.4}SrCoO_x$ which yields cationic compositions '$Pb_{1/3}SrCo$' after reaction.

The structure of this misfit [13] is isostructural to that of $Tl_{0.4}SrCoO_{3+\delta}$ and $Ca_3Co_4O_9$, i.e. three layers [SrO]-[(Pb/Sr)O]-[SrO] are interleaved with the $[CoO_2]$ layer.

One interesting property of the Tl-, Bi/Pb- and Ca-based misfit cobaltite lies in their large thermopower (S) at room temperature, always in the range $90\mu V.K^{-1}$ to $130\mu V.K^{-1}$, which is similar to $S^{300K} = +100\mu V.K^{-1}$ reported for $NaCo_2O_4$ [1]. From all these thermopower results, it is tempting to associate the large S value with the peculiar $CoO_2$ layer common to all these oxides. The thermopower values obtained for the pure lead misfit cobaltite reinforce this hypothesis. As shown in Fig. 5, at RT the value for the Pb/Sr/Co/O misfit is found to be close to $+110\mu V.K^{-1}$. This value is also close to that exhibited by $(Tl/Bi)_{0.4}SrCoO_{3+\delta}$ in Fig. 4. And similarly, the resistivity at low T exhibits also a reentrance (Fig. 5). The T dependent ρ curve exhibits thus a ρ minimum at about $T_{min}$ = 90K for the pure Pb cobaltites.

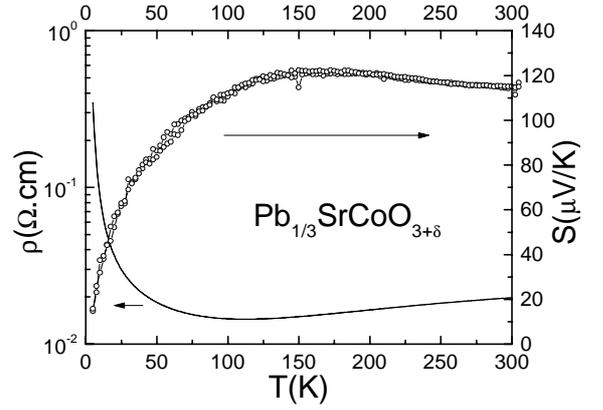

Fig. 5 : S(T) and ρ(T) curves of $Pb_{1/3}SrCoO_{3+\delta}$

The physical properties of these new Pb misfit cobaltite are also very like that of $Ca_3Co_4O_9$ which exhibits same kind of S(T) and ρ(T) characteristics. A negative MR is also found in $Ca_3Co_4O_9$.

## 2.3. CONCLUSIONS

Among the misfit cobaltites, the Tl/Sr/Co/O one is unique since this is the only compound which exhibits metallicity and paramagnetism down to the lowest temperature accompanied by a positive magnetoresistance [12]. As discussed in the paper by Singh on the electronic structure of $NaCo_2O_4$ [8], it is very difficult to predict the low T magnetic state of the cobalt oxides made of $CdI_2$-type $CoO_2$ layers since magnetically ordered or paramagnetic solutions cannot be distinguished. In this respect, the chemical route consisting in the Bi for Tl substitution is interesting since on the one hand, it induces at low T, a re-entrant resistivity and negative magnetoresistance, and, on the other hand, an increase of the thermopower value at RT. In this respect, the new Pb/Sr/Co/O misfit behaves more like the Tl/Bi-misfit cobaltite than the pure Tl one. It should be pointed out that in these cobaltites, all the compounds exhibiting a RT thermopower value larger than that of Tl-based misfit show a negative magnetoresistance linked to the re-entrance in the resistivity at low T. This demonstrate the existence of the interplay between charges and spins in this class of oxides.

Since all the thermopower values of misfit cobaltites are close to those of $NaCo_2O_4$, one has to consider that the peculiar rhomboedral splitting of the $t_{2g}$ orbitals of $Co^{3+}/Co^{4+}$ of the $CoO_2$ layer into $a_{1g}$ and $e'_g$ levels is responsible for these high values. In this framework, it was proposed that the narrow band made of heavy holes in the $a_{1g}$ orbitals mostly contributes at the Fermi level to yield large thermopower [8]. It appears thus that this band is not very sensitive to the nature of the separating layers interleaved between the $CoO_2$ layer. In contrast, the different behaviors of resistivity and magnetism can be interpreted as the result of a modification in the filling or in the position of the $e'_g$ band made of mobile holes. However, the composite nature of these misfit oxides makes very difficult the refinements of their structures. Subtle changes in the cell parameters can affect the $a_{1g}$ and $e'_g$ respective contributions at $E_F$. Single-crystals are now required to quantify the mobile holes density by Hall effect and to check simultaneously the structural changes induced by Bi for Tl partial substitution or the complete replacement of Tl by Hg. Thermal conductivity measurements are also required to compare the figures of merit of these different misfit cobaltites to that of $NaCo_2O_4$.

## 3. MANGANESE OXIDES
### 3.1 CMR PROPERTIES

Manganese oxides have been the subject of a renewed attention in the last few years after the discovery of Colossal Magneto Resistance (CMR) [for recent reviews see ref. 14].

Manganites crystallize in the perovskite structure. It can be described as a cubic close packed array with O anions at the centers of the faces, the lanthanide and alkaline rare-earth (La, Pr, Ca, Sr…) at the corner (site A) and Mn ion at the center in the octahedral interstitial site (site B). The ideal cubic structure is strongly distorted due to the cation size mismatch and due to the Jahn-Teller effect active for $Mn^{3+}$.

The parent compounds (for example, $LaMnO_3$ with a manganese valency equal to 3 or $CaMnO_3$ with only $Mn^{4+}$) are anti-ferromagnetic insulators. The $Mn^{3+}$ and $Mn^{4+}$ electronic structures are respectively $3d^4$ and $3d^3$, which following Hund's rule correspond to orbital filling $t_{2g}^3 e_g^1$ and $t_{2g}^3$. When doping the A site with a divalent cation (for $LaMnO_3$) or trivalent cation (for $CaMnO_3$), mixed valent manganese is introduced and CMR properties can appear. They are linked to the appearance of ferromagnetism in these materials, usually simultaneously occurring with a metal/insulator transition separating the high temperature paramagnetic insulating phase from the ferromagnetic metallic low temperature phase. For the optimal composition corresponding to 30% of $Mn^{3+}$, a huge magnetoresistance is observed as for example in $Pr_{0.7}Ca_{0.26}Sr_{0.04}MnO_3$ [15] where the resistance drops by more than 11 orders of magnitude when applying 5T.

A first attempt to explain the CMR properties is based on the double-exchange model [16]. In such a model, the hopping integral of the $e_g$ electron is proportional to the angle between the manganese ions $Mn^{3+}/Mn^{4+}$. When applying a magnetic field, the alignment between manganese ions is favored and the resistivity decreases, leading to negative magneto-resistance. It was however shown that double-exchange model can qualitatively but not quantitatively explain CMR [17] and several studies have revealed the importance of electron-phonon interactions to understand these phenomena. These manganites are actually a nice framework to investigate the interplay between spins, phonons and carriers.

As CMR is more important for the hole doped manganites than for the electron doped ones, most of the studies have focused on this doping side, with investigation of the A site or B site doping to try to optimize CMR and to understand its origin. More recently, it was found that CMR can also been found in electron doped manganites such as $Ca_{1-x}Sm_xMnO_3$ with x~0.15[18].

In both cases (electron or hole doped manganites) thermoelectric properties, as other transport and magnetic properties, are dramatically modified by doping. The undoped insulating compounds have in both cases large thermopower and high resistivity : for example at room temperature, S=-350µV/K and ρ=2Ωcm for $CaMnO_3$, and S =+450µV/K and ρ=10Ωcm for $LaMnO_3$. With doping, thermopower decreases as holes or electrons are introduced. Simultaneously resistivity is decreased and because of the lower concentration of $Mn^{3+}$ Jahn-Teller cations for the electron side, resistivity can be as low as 1mΩcm (see [6]), much smaller than the values typically of the order of a few Ωcm obtained for doped $LaMnO_3$. For thermoelectric properties, a large $Z=S^2/(\kappa\rho)$ is required and the electron doped manganites are therefore better candidates for thermoelectric applications.

We will show that by doping $CaMnO_3$, it is possible to get a low resistivity while keeping a large S, thus reaching a reasonable Z, similar to the highest Z measured in oxides ($NaCo_2O_4$ [1] or $La_xSr_{1-x}TiO_3$ [2]). We will describe in the following the evolution of thermopower when dopant is introduced on the A or B site of $CaMnO_3$ and will focus on the efficiency of Ru when substituted on the B site.

### 3.2 A SITE DOPING OF $CaMnO_3$

Results concerning the series $Ca_{1-x}Sm_xMnO_3$ (x≤ 0.12) are presented on Fig. 6.

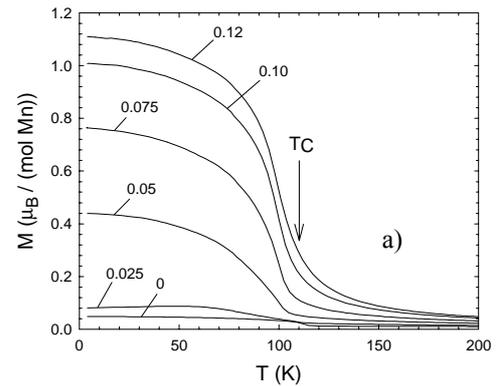

Fig. 6 : a) M(T) curves measured under an applied field of 1.45T for $Ca_{1-x}Sm_xMnO_3$.

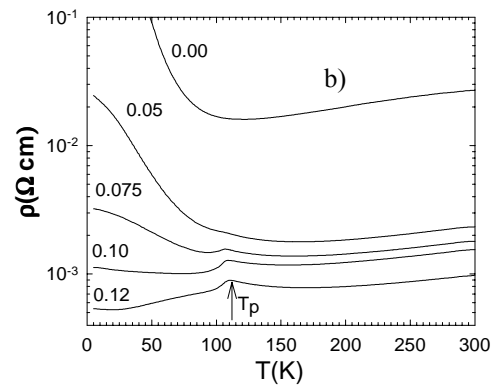

b) ρ(T) curves measured under zero applied field

The magnetization (fig. 6a) was measured under 1.45T after zero field cooling as a function of temperature. The substitution of Ca by Sm leads to ferromagnetic interactions but the magnetic moment remains smaller than the expected saturation value (1.1 $\mu_B$ for x=0.12 at 5K, compared to 3.12$\mu_B$). AC susceptibility measurements have revealed a complex behavior of 'cluster glass' at low temperatures [6] corresponding to short-range ferromagnetic clusters. In figure 6b, $\rho(T)$ curves are shown, measured under 0T. It is evident that introducing electrons in the $Mn^{4+}$ rich matrix has beneficial effects on resistivity, as it is decreased in the whole temperature range as x increases. Note that even $CaMnO_3$ presents here a metallic behaviour for T>100K with low resistivity which is induced by a small oxygen deficiency ($CaMnO_{3-\epsilon}$ with $\epsilon \approx 0.02$ [6]) creating carriers. For stoechiometric $CaMnO_3$, resistivity is much larger, reaching 2$\Omega$cm at 300K [19]. As x increases, $\rho$ values as low as 1m$\Omega$cm are obtained at 300K for x=0.12. The high temperature resistivity between 200K and 300K can be fitted by a small polaron model with

$$\rho = AT\exp(W/kT)$$

where A is constant and W is the hopping energy (varying from 5 to 10 meV when x increases from 0 to 0.12). The fact that the polaron energy increases even if the metallicity increases probably reflects the growing concentration of $Mn^{3+}$ which are sensitive to Jahn-Teller effect and favor the formation of polarons. The peak in $\rho(T)$ observed around 110K corresponds to the Curie temperature Tc (Fig. 6a). At low temperatures, the resistivity behavior is evolving from semi-conducting for small x to metallic for x> 0.05. The most important point here is that room temperature resistivity as low as 1m$\Omega$cm can be obtained.

Figure 7 presents the thermopower of the different studied samples.

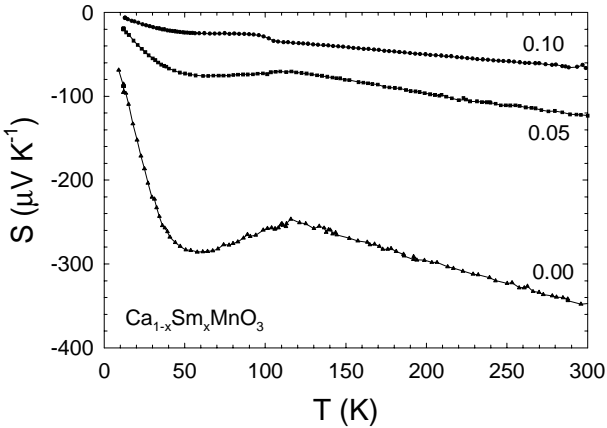

Fig. 7 : S(T) of the compounds $Ca_{1-x}Sm_xMnO_3$.

As x increases, the metallicity increases which induces a decrease of the absolute value of thermopower. The values of S at 300K remain nevertheless large for metals (for example, -120$\mu$V/K for x=0.05). Thermopower is negative in the whole temperature range with a linear temperature dependence in the paramagnetic region. Below Tc, an increase of |S| (for x=0.05 and x=0) or a decrease (for x=0.1) associated with the onset of magnetic ordering is observed, then the thermopower decreases to very small values as T tends to 0. S can be described by a single band metal model with

$$S = \pi^2 \times k_B/3e \times k_B T (\partial \ln \sigma(E)/\partial E).$$

Using the values of S and $\sigma$ at 300K, the number of carriers n(E) (with $\sigma(E)=n(E)\times e \times \mu(E)$, where $\mu(E)$ is the carrier mobility) can be extracted and is found to be very close to the values expected from chemical formula (see [6]) with typically N(E)/n(E) close to 7-12 depending on x (with N(E) the density of states).

Combining S and $\rho$ values at room temperature, the power factor P=$S^2/\rho$ is equal to 7$\mu$WK$^{-2}$cm$^{-1}$ for x=0.05, a value very large compared to other oxides (except $NaCo_2O_4$ and cobalt misfits) and only 4 times smaller than the power factor of $Bi_2Te_3$ [1]. This value is still smaller than the best reported power factor obtained in single crystals of $NaCo_2O_4$ of 50$\mu$WK$^{-2}$cm$^{-1}$ [1]. Note however that the properties reported here concern polycrystalline samples for which conductivity could be improved and thus increase power factor. In the case of $NaCo_2O_4$ there is indeed a large difference between the power factor of crystals (50$\mu$WK$^{-2}$cm$^{-1}$ [1]) and the one of polycrystalline samples (2$\mu$WK$^{-2}$cm$^{-1}$ [20]).

To calculate the figure of merit Z = $S^2/(\rho\kappa)$, thermal conductivity measurements have been carried out. Figure 8 presents the results obtained for the same compounds as those of Figure 7.

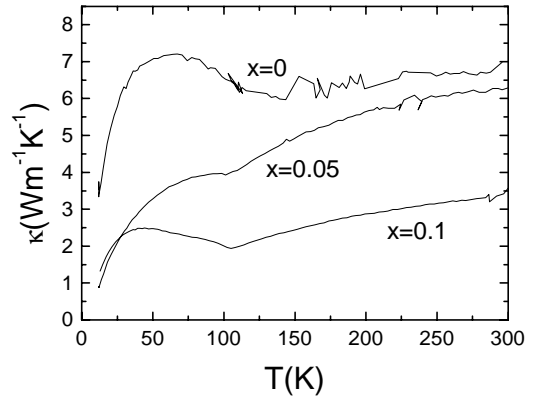

Fig. 8 : $\kappa$(T) of the compounds $Ca_{1-x}Sm_xMnO_3$

The thermal conductivity is small for these metallic oxides (maximum 7Wm$^{-1}$K$^{-1}$), but still much larger than for $NaCo_2O_4$ for which a value of 2Wm$^{-1}$K$^{-1}$ was reported at room temperature [20]. This relatively large value is not favorable to get large Z. A reasonably small value of $\kappa$ can be reached only for large x but as increasing x in $Ca_{1-x}Sm_xMnO_3$ would only decrease S and increase $\rho$ [21], another route to improve thermoelectric properties has been investigated.

### 3.3 Mn SITE DOPING IN CaMnO$_3$

Several cations have been investigated as possible dopants on the B site of $CaMnO_3$. First studies were dedicated to the investigation of CMR properties and the idea was to substitute Mn by a cation with a valency superior to 4 to induce mixed valency $Mn^{3+}/Mn^{4+}$ and thus favor CMR properties.

Using non magnetic cations $d^0$ (Nb, Ta, W and Mo) as susbtitution cation M in $CaMn_{1-x}M_xO_3$, it was indeed found that CMR appears for x as small as 0.02 [19]. This is due only to a valency effect which favors the appearance of mixed valency together with double-exchange interactions, and disappears for tetravalent cations such as $Ti^{4+}$.

To further improve CMR, more attention has been paid to Ru because of its possible mixed valent state $Ru^{4+}/Ru^{5+}$ with isoelectronic structure of $Mn^{3+}/Mn^{4+}$ ($d^4$ and $d^3$) which could enhance ferromagnetic interactions by super-exchange coupling between Ru and Mn.

Figure 9 represents the magnetization as a function of temperature measured under 1.45T after zero-field cooling for $CaMn_{1-x}Ru_xO_3$ ($0 \le x \le 0.5$).

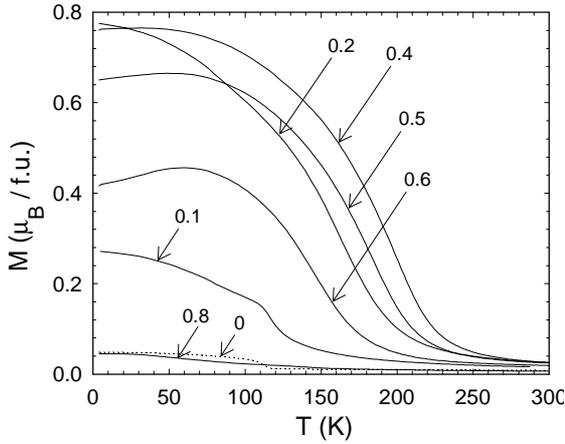

Fig. 9 : M(T) curves measured under an applied field of 1.45T of $CaMn_{1-x}Ru_xO_3$.

In a similar manner as in $Ca_{1-x}Sm_xMnO_3$, magnetic moment continuously increases when x increases, reaching $0.8\mu_B$ for x=0.4 and then decreases again for x>0.4. Ferromagnetism is induced by Ru doping but the magnetic moment remains very small compared to the expected moment as for A site doping. This reveals a non uniform material with phase separation phenomena : at room temperature, the material is structurally single-phase (orthorhombic), but at low temperatures, high resolution electronic microscopy performed at T=92K reveals the existence of an antiferromagnetic distorted phase coexisting with ferromagnetic orthorhombic regions. Most interestingly, this phase separation phenomenon leads to a large increase of magneto-resistance [22].

Resistivity measurements are presented in Fig. 10. $\rho$ continuously decreases as x increases up to x=0.12. For larger x, the room temperature resistivity increases while the low temperature values remain small. For a thermoelectric use close to room temperature, the range of possible x is therefore limited to x < 0.2. A minimum value of $\rho = 2m\Omega.cm$ is obtained at 300K for x=0.12. As for the series $Ca_{1-x}Sm_xMnO_3$, these data can be fitted by the small polaron model, but contrary to this case, the hopping energy decreases as x increases (from 37.5meV for x=0.06 to 28.4 meV for x=0.1) [23].

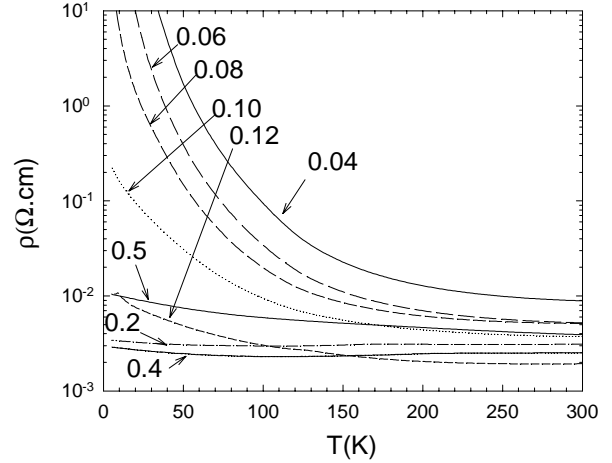

Fig. 10 : $\rho(T)$ curves measured under zero applied field of $CaMn_{1-x}Ru_xO_3$.

Compared to the other non magnetic cations previously used (Nb, Mo…), $\rho^{300K}$ is smaller in the case of Ru than in the case of non magnetic cation. For example, for Nb, a minimum value of 8 m$\Omega$.cm is obtained for 6% of Nb. In the case of non magnetic cations, even if ferromagnetism is induced and favors metallicity, the presence of foreign cations on the B site disturbs the hopping transport of electrons along the Mn network. On the other hand, for Ru, ferromagnetic interactions and thus metallicity are enhanced because of superexchange interactions between Mn and Ru and furthermore, the $e_g$ electron of $Ru^{4+}$ can participate in the transport phenomena, reducing the 'electronic disorder' on the Mn network [24].

Thermoelectric properties of the Ru doped $CaMnO_3$ are presented in Fig. 11. As metallicity is enhanced when Ru is introduced, the TEP S is decreased in the whole temperature range compared to the undoped compound $CaMnO_3$. For a x value which corresponds to reasonably small values of $\rho$ (5m$\Omega$.cm for x=0.06), S at 300K is decreased but remains very large close to -140$\mu$V/K.

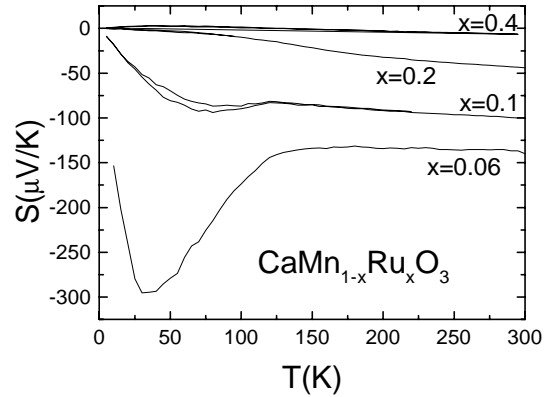

Fig. 11 : S(T) of the compounds $CaMn_{1-x}Ru_xO_3$.

Larger values of x induce a strong decrease of S together with a stong increase of $\rho$.

Ru is a unique dopant for $CaMnO_3$ as its electronic structure similar to that of Mn favors the interactions between Mn and Ru and reinforces metallicity. Furthermore, the other advantage, compared to other dopants, is that because of its mixed valency (between 4 and 5), it decreases the Mn valency less rapidly than other pentavalent or hexavalent cations. S is very sensitive at high temperatures to the carrier concentration and following Heikes formula, decreases as the Mn valency decreases. The slow decrease of Mn valency induced by Ru thus preserves large values of S compared to other dopants. This is illustrated in Fig. 12.

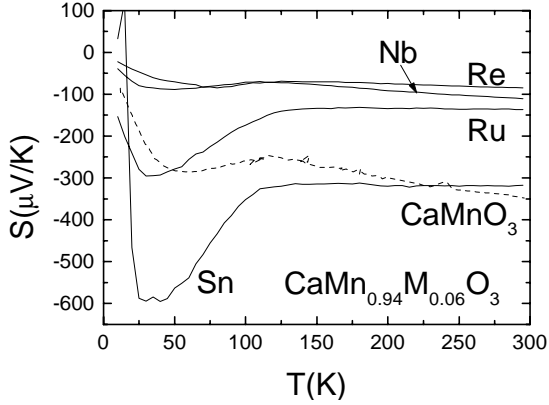

Fig. 12 : S(T) of $CaMn_{0.94}M_{0.06}O_3$

Except for tetravalent $Sn^{4+}$ (which does not induce metallicity), S is strongly reduced by the substitution of 6% of foreign cation. However, the reduction is less drastic for Ru than for Nb or Re, both pentavalent cations. At room temperatures, even if the three compounds have similar resistivity, the power factor will be larger for Ru.

Figure 13 presents the thermal conductivity of the same compounds as in Fig. 12.

Fig. 13 : $\kappa(T)$ of $CaMn_{0.94}M_{0.06}O_3$

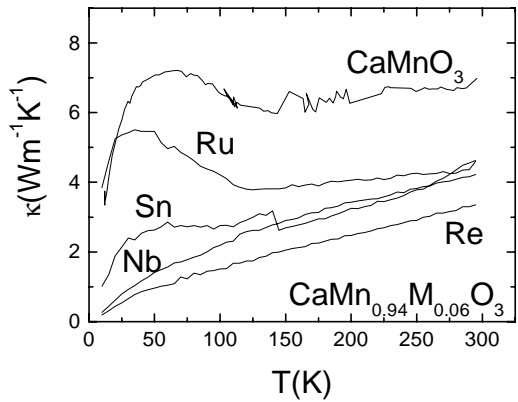

Thermal conductivity is rather small as for $Ca_{1-x}Sm_xMnO_3$ but the reduction of $\kappa$ at room temperature is more important than for the A site doping. $\kappa$ remains smaller than $5 Wm^{-1}K^{-1}$ for 6% of Ru. Considering the Lorenz number $L=\kappa/\sigma T = 2.44 \ 10^{-8} \ W\Omega K^{-2}$, the estimated electronic contribution to $\kappa$ is close to $0.24 Wm^{-1}K^{-1}$ at room temperature in the case of Ru, showing that $\kappa$ is dominated by the phonon contribution. Combining $\rho$, S and $\kappa$, we obtain a figure of merit equal to $1.33 \ 10^{-4} \ K^{-1}$ at 300K which makes it as good as the best thermoelectric oxide $NaCo_2O_4$.

The stronger reduction of $\kappa$ in the case of B site doping compared to $Ca_{1-x}Sm_xMnO_3$ is linked most probably to the fact that disorder on the B site is detrimental for the heat propagation as carriers and phonons are strongly coupled, as evidenced by the small polaron behavior of resistivity.

### 3.4 CONCLUSIONS

Table 1 summarizes the results obtained at room temperature with manganese oxides and compare them to the other best thermoelectric oxides and to 'standard' thermoelectric materials.

|  | Ru doped $CaMnO_3$ x=0.06 | Sm doped $CaMnO_3$ x=0.05 | $NaCo_2O_4$ | $La_{1-x}Sr_xTiO_3$ (x=0.05 [2]) | $Bi_2Te_3$ |
|---|---|---|---|---|---|
| S (μV/K) | -140 | -120 | +100 [1] +80 [20] | -150 | 200 [1] |
| ρ (mΩcm) | 5 | 2 | 0.2 [1] 3 [20] | 0.5 | 1 [1] |
| κ ($Wm^{-1}K^{-1}$) | 4 | 6 | 2 [20] | 12 | |
| $P=S^2/\rho$ (μW$K^{-2}cm^{-1}$) | 4 | 7 | 50 [1] 2 [20] | 45 | 40 [1] |
| ZT at 300K | 0.030 | 0.036 | 0.032 [20] | 0.110 | 1.2 [25] |

Table 1 : Comparison of the thermoelectric performances of the different oxides studied here with the best thermoelectric materials.

The manganese oxides investigated here have similar properties as the other thermoelectric oxides. By chosing the right dopant, the different factors of Z can be tuned to optimize it : Ru is promising as it introduces disorder among the Mn ions and reduces $\kappa$, it favors metallicity because of its electronic structure, and last the reduction of S linked to the improved metallicity is not too large at room temperature because of its mixed valency. Furthermore, as in the case of $NaCo_2O_4$, a large reduction of $\rho$ is expected in the case of single crystals which could strongly improve Z. The negative thermopower makes them interesting as n leg for a thermoelectric device. Indeed, it has been very recently reported that $Ca_{0.92}La_{0.08}MnO_3$ has already been used to design a thermoelectric power generator with promising performances [3].

**REFERENCES**
[1] : I. Terasaki, Y. Sasago, K. Uchinokura, Phys. Rev. B 56, R12685 (1997).
[2] : T. Okuda, K. Nakanishi, S. Miyasaka and Y. Tokura, Phys. Rev. B 63, 113104 (2001).
[3] : I. Matsubara, R. Funahashi, T. Takeuchi, S. Sodeoka, T. Shimizu, K. Ueno, Appl. Phys. Lett. 78, 3627 (2001).

[4] : Ph. Boullay, B. Domengès, M. Hervieu, D. Groult and B. Raveau, Chem. Mater. 8, 1482 (1996).
[5] : H. Leligny, D. Grebille, O. Perez, A.C. Masset, M. Hervieu, C. Michel and B. Raveau, C. R. Acad. Sci., Paris, Série Iic2, 409 (1999).
[6] : A. Maignan, C. Martin, F. Damay, B. Raveau, J. Hejtmanek, Phys. Rev. B 58, 2758 (1998).
[7] : M. Von Jansen and R. Hoppe, Z. Anorg. Allg. Chem. 408, 104 (1974).
[8] : D. J. Singh, Phys. Rev. B 61, 13397 (2000).
[9] : M. A. Senaris-Rodriguez and J. B. Goodenough, J. Solid State Chem. 118, 323 (1995).
[10] : T. Yamamoto, Ph. D. Thesis, University of Tokyo, 2001.
[11] : A.C. Masset, C. Michel, A. Maignan, M. Hervieu, O. Toulemonde, F. Studer, B. Raveau, and J. Hejtmanek, Phys. Rev. B 62, 166 (2000).
[12] : S. Hébert, S. Lambert, D. Pelloquin and A. Maignan, to be published in Phys. Rev. B 64, 1321xx.
[13] : (unpublished).
[14] : C. N. R. Rao, B. Raveau (Eds.), Colossal Magnetoresistance, Charge ordering and Related Properties of Manganese Oxides, World, Scientific, Singapore 1998 ; Y. Tokura (Ed.), Colossal Magnetoresistive Oxides, Gordon and Breach Science Publishers, New York 1999.
[15] : A. Maignan, Ch. Simon, V. Caignaert and B. Raveau, Solid State Comm. 96, 623 (1995).
[16] : C. Zener, Phys. Rev. 81, 440 (1951).
[17] : A. J. Millis, P. B. Littlewood, B. I. Shraiman, Phys. Rev. Lett. 74, 5144 (1995).
[18] : C. Martin, A. Maignan, F. Damay, M. Hervieu, B. Raveau, J. Solid State Chem. 134, 198 (1997).
[19] : B. Raveau, Y. M . Zhao, C. Martin, M. Hervieu, A. Maignan, J. Solid State Chem. 149, 203 (2000).
[20] : I. Terasaki, Proc. 18[th] Int. Conf. Thermoelectrics (ICT'99) Baltimore, 1999 (IEEE, Piscataway 2000), p 569.
[21] : J. Hejtmanek, Z. Jirak, M. Marysko, C. Martin, A. Maignan, M. Hervieu, B. Raveau, Phys. Rev. B 60, 14057 (1999).
[22] : A. Maignan, C. Martin, M. Hervieu, B. Raveau, Solid Stat. Comm. 117, 377 (2001).
[23] : B. Raveau, A. Maignan, C. Martin, M. Hervieu, Mater. Res. Bull. 35, 1579 (2000).
[24] : C. Martin, A. Maignan, M. Hervieu, C. Autret, B. Raveau, D. I. Khomskii, Phys. Rev. B 63, 174402 (2001).
[25] : G. D. Mahan, Solid State Phys. 51, 81 (1998).